\documentstyle[11pt,paspconf,twoside,epsf]{article}
\def\edcomment#1{\iffalse\marginpar{\raggedright\sl#1\/}\else\relax\fi}
\reversemarginpar

\begin{document}
\title{The Intergalactic Medium and the Energy Distribution of 
Quasars }

\author{Luc Binette, Mario Rodr\'\i guez-Mart\'\i nez and Isidro
Ballinas}
\affil{Instituto de Astronomía, Universidad Nacional Aut\'onoma de
M\'exico, Apartado Postal 70-264, 04510 M\'exico, 
DF, Mexico}

\begin{abstract}
The ionizing spectral energy distribution of quasars is known to
exhibit a steepening of the distribution short-ward of 1050\AA. It has
been assumed that this change of power-law index from $-1$ to $-2$ is
intrinsic to the quasars. We study an alternative interpretation, in
which a tenuous absorption screen is responsible for the change of
index.  A successful spectral fit is obtained using a total hydrogen
density distribution decreasing with redshift as $\sim (1+z)^{\xi}$
with $\xi$ of the order $-3.5$ if we suppose that the ionization state
of the intervening gas is controlled by the line-of-sight quasar. This
interpretation is not compatible, however, with the relative weakness
of the observed quasar intensities with respect to that of the local
background metagalactic radiation. Therefore, our interpretation in
its present form cannot be sustained.
\end{abstract}

\section{Introduction}

The ionizing spectral energy distribution (hereafter ISED) of nearby
active galactic nuclei cannot be observed directly due to the galactic
absorption beyond the Lyman limit. Owing to the redshift effect,
however, we can get a glimpse of the ISED from the spectra of very distant
quasars. Most studies on the subject indicate a steepening of the
distribution at wavelengths shorter than $~ 1000\,$\AA\ (see O'Brien
et~al. 1988) and references therein).  The most extensive work on the
subject has been performed by Zheng et~al. (1997, ZKTGD) using HST
archived data.  They found that the power-law index ($F_{\nu}\propto
\nu^{\alpha}$) steepens from $\approx -1$ for $\lambda > 1050\,$\AA\ to 
$\approx -2$ for shorter wavelengths. 
 
Korista, Ferland \& Baldwin (1997) pointed out that such a steep slope
for the ISED would imply an insufficient number of photons beyond $h\nu
> 54.4\,$eV to account for the observed luminosities of the high
excitation emission lines.

We report preliminary results of our attempt to find an alternative
explanation for the break, one that is {\em extrinsic} to the quasar
ISEDs. We postulate the existence of a very tenuous absorption gas
component pervading the local universe and proceed to study the
characteristics it must possess to reproduce the observed break, and
its place and role within our IGM views.

\section{Data simulation to a first order}  

The composite spectrum of ZKTGD was constructed by merging 284 spectra
of 101 quasars taken with FOS of the Hubble telescope. Although the
wavelength coverage of individual spectra was relatively narrow, the
distribution in quasar redshifts allowed ZKTGD to cover the wavelength
range of 310--3000\,\AA\ in the quasar rest-frame. Before merging
their spectra, each spectrum was corrected for the Lyman valley and
Ly$\alpha$ forest absorption by calculating the appropriate
transmission curve using the scheme developed by M\o ller \& Jakobsen
(1990).

Our aim is to reproduce the change of ISED index by introducing a
previously unknown intergalactic absorption component. This component
must consist of clumps which are sufficiently optically thin (and
numerous) that they do not show up as individually detectable Ly$\alpha$
absorption lines in the archived FOS spectra, hence $N_{H^0} \ll
10^{12}\,{\rm cm^{-2}}$. 
In this preliminary work, we did not fully simulate the
observed data on a pixel by pixel basis, rather we defined a single
spectrograph wavelength window (3000\,\AA\ to 1300\,\AA) which we
considered typical of the wavelength coverage of individual quasars
and then redshifted this window in locked steps. We defined a {\it
source} array, $F_j$, to represent the intrinsic SED of all
quasars. This SED consisted of a power-law of index $\alpha$. All
arrays were evenly binned in $Log \lambda$ and the wavelength of each
array element $j$ was such that $Log \lambda_j = Log
\lambda_{MAX} - j \times u$, with $u = Log (1+\delta z_{o})$ and
$\lambda_{MAX}$ being the highest rest-wavelength considered.  The
advantage of this scheme is that the source array $F_j$ can be
redshifted by simple translation of the pointers $j$ provided the
redshift is a multiple of $\delta z_{o}$. This is not a serious
limitation if one chooses a sufficiently small $\delta z_{o}$ (we used
$0.01$). The procedure followed consisted in translating the source
array by an integer number of bins $k$ and then add it to our
averaging buffer: $B_j = B_j + w_k F_j T_j(z_Q^k,\lambda_j)$ where
$T_j$ is the transmission function  evaluated
at $\lambda_j$ for the $k^{th}$ array spectrum (corresponding to a
quasar's redshift $z_Q^k=k \,
\delta z_{o}$), and $w_k$ is the global weight given to the $k^{th}$
array before summing. These `uniform' weights were based on the
dispersion vs. rest-frame wavelength relation displayed in the Fig.~6
of ZKTGD. More specifically, $w_k = s^{-2}_k$ where $s_k =
\bar\lambda_{kc}^{-0.71}$ with $\bar\lambda_{kc} = 2150 (1+z_Q^k)$\,\AA,  the
central wavelength of redshifted array $k$.  The lowest and highest
redshifts considered were 0.33 and 3.5, respectively, which are the
values quoted by ZKTGD for their data set.  We uniformly covered this
redshift interval by considering all integer values of $k$ between
these two redshift limits.

\section{The transmission function}  

For each quasar rest-frame wavelength bin $j$, 
we calculated the transmitted intensity
$I_{\lambda_j}^{tr} = I_{\lambda_j} T_{\lambda_j} = I_{\lambda_j} 
e^{-\tau(\lambda_j)}$ by integrating the
opacity along the line-of-sight to a quasar of redshift $z_Q$ 
\begin{equation}
\begin{array}{cc}
\tau(\lambda_j) = \sum_{i=0}^{10}{\int_{0}^{z_Q}
\sigma_i(\frac{\lambda_j}{1+z})\; n_{H^0}(z) \frac{dl}{dz} \; dz}  \label{eq:tau}
\end{array}
\end{equation}
where $\lambda_j$ is the quasar rest-frame wavelength for bin $j$ and
$n_{H^0}(z)$ the intergalactic neutral hydrogen density. The summation
was carried out over the different opacity sources: photoionization
($i=0$) and line absorption from the Lyman series of hydrogen ($1\le i
\le 10$). Although our code could include up to 40 levels, we found
that considering only the 10 lowest proved to be adequate.  We adopted
a fiducial velocity dispersion $b$ of 30~km/s and assumed a simple
line Gaussian profile for $\sigma_i$ of the lines.  Apart from the
uniform density absorption case, we also implemented the treatment of
a clumpy medium (described later) which was used to ensure that we
could reproduce the transmission curve of M\o ller \& Jakobsen (1990)
or ZKTGD using parameters appropriate to the Ly$\alpha$ forest. The self
imposed constraint that $N_{H^0} \ll 10^{12}\, {\rm cm^{-2}}$ for the clumps
ensures that we remain in the linear regime of the curve of
growth. Therefore, using a uniformly distributed gas or a clumpy
medium will be equivalent. The use initially of an homogeneously
distributed gas will save us from having to characterize the
clumpiness using poorly defined parameters.

For the dependence of the total gas density on redshift, let us
consider a simple power-law $n_H \propto (1+z)^{\delta}$.  What is
lacking is the neutral fraction of $n_H$ which we will assume to be small
and set by equilibrium between photoionization and recombination. In
that case, we expect $n_{H^0}$ to scale proportionally to the (total)
density square, and inversely proportionally to the ambient ionizing
radiation. We will consider negligible the cumulative opacity of the
absorber (as confirmed a posteriori).  With those considerations in
mind, we obtain that $n_{H^0}$ should scale as follows with $z$:

\begin{equation}
\begin{array}{cc}
 n_{H^0}(z) = n^0_{H^0} {(1+z)^{2(\xi+3)}} \, \Gamma^{-1}(z) \,
{(1+\omega)^{-1}} \label{eq:nh}
\end{array}
\end{equation}
where the upperscript {\small 0} throughout the text denotes
quantities evaluated at $z=0$.  In the above expression, we take into
account the increase of density with the cosmological scale factor:
$n_H/n_{H}^0 \propto (1+z)^3$. The photoionization rate coefficient
$\Gamma$ due to the metagalactic ionizing radiation was parameterized
as follows
\begin{equation}
\begin{array}{c}
\Gamma(z) = (1+z)^{0.73} e^{-0.526 ((z - 2.3)^2 - 2.3^2))} \label{eq:madau}
\end{array}
\end{equation}
This is a renormalized version of the parametric fit of
Haardt \& Madau (1996) who modeled the metagalactic ionizing radiation
($J_{\nu_0}^{BCK}$) taking into account the population evolution of
quasars and their luminosity with redshift as well as self-emission of
the clumpy intergalactic ionized medium. Note that $\Gamma(z=0) = 1$
because of our renormalization.

The factor $(1+\omega)^{-1}$ in equation~2 is the same as that defined
by Bajtlik, Duncan \& Ostriker (1988) and represents a decrease in the
neutral fraction of hydrogen towards the line-of-sight quasar as a
result of the ionizing radiation from the quasar.  In this expression,
the ``proximity'' term $\omega$ represents the geometrical dilution of
the line-of-sight quasar ionizing flux with respect to the background
ionizing radiation:
\begin{equation}
\begin{array}{cc}
 \omega = \left(\frac{r_{1/2}}{r_L}\right)^2 \label{eq:omeg} 
\end{array}
\end{equation}
where $r_L$ is the luminosity distance of the quasar from a parcel of
gas at $z$.  $r_{1/2}$ if one of our free parameters and correspond to
the distance from the quasar where its mean intensity at $\nu_0$
equals that of the background radiation $J_{\nu_0}^{BCK}$. At such
distance, the neutral fraction $x=n_{H^0}/n_H$ is halved due to the
quasar flux. $r_{1/2}$ can be expressed as follows
\begin{equation}
\begin{array}{cc}
r_{1/2} = r_{1/2}^0 \sqrt{ (1+z_Q^{\prime})^{\alpha -1} \, \Gamma^{-1}(z)}  \label{eq:rmedz}
\end{array}
\end{equation}
with
\begin{equation}
\begin{array}{cc}
r_{1/2}^0 =  \frac{1}{4\pi} \sqrt{ {{L_{\nu_0}^Q}\over{J_{\nu_0}^{BCK,0}}} }\label{eq:rmedi}
\end{array}
\end{equation}
where $\alpha$ is the power-law index of the quasar spectral
luminosity distribution ($L^Q_{\nu} \propto \nu^{\alpha}$) and $\nu_0$
is the hydrogen photoionization threshold frequency.  As ISED index,
we adopt $\alpha = -0.86$ which is the average index for the
radio-quiet subsample of ZKTGD (above 1050\AA).  In our data simulation,
to describe the behavior of $n_{H^0}$ with redshift, we employ
only two free parameters: $\xi$ and $r_{1/2}^0$.

\begin{figure}
\plotone{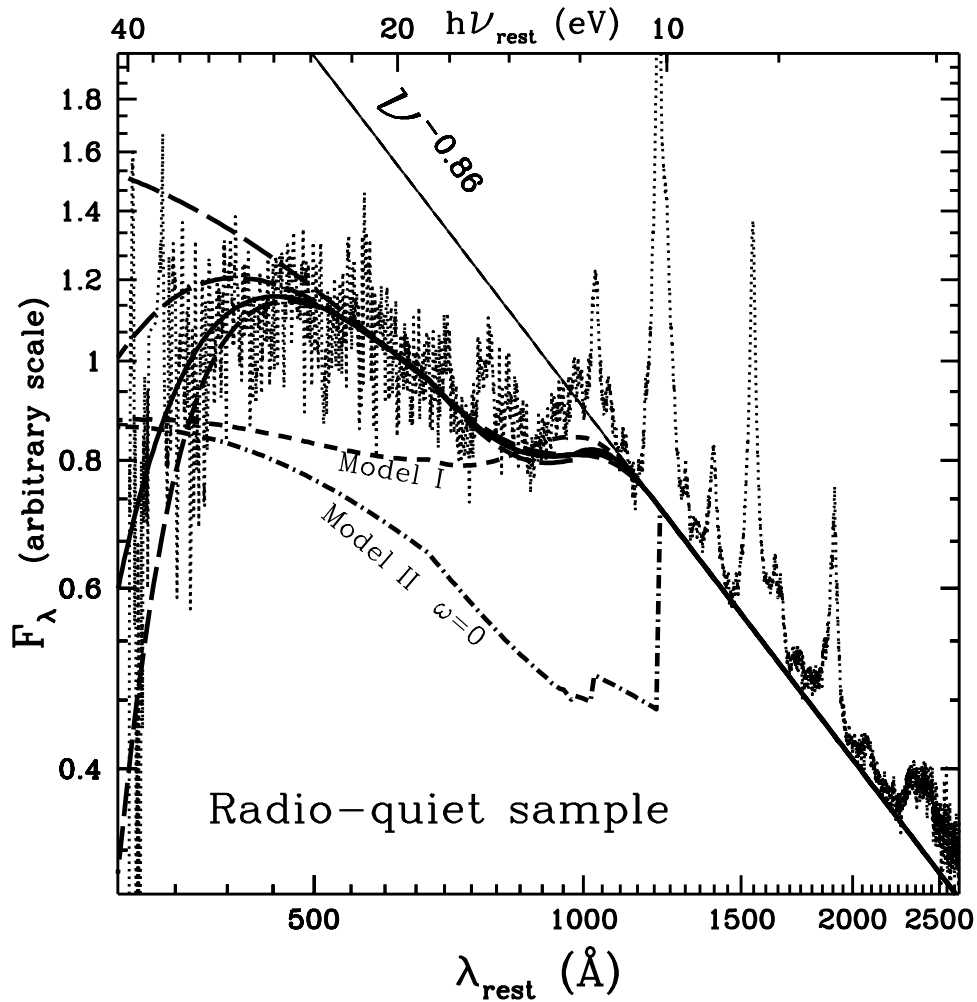}
\caption{The composite spectrum of radio-quiet quasars constructed by
Zheng et~al. (1997) (dotted line). The straight short-dashed line  
 represents a power-law fit ($\alpha = -0.86$) above $1050$\AA\ to the
 composite continuum underlying the emission lines.  The solid line is
 our favored Model~IV ($\xi = -4$, see Table~1).  Three
 other models using different values of $\xi$ are represented by the
 long-dashed lines. From the top to the bottom curve, $\xi$ is
 decreasing from $-2$ to $-5$. The horizontal short-dashed line is
 Model~I which does {\it not} include redshift evolution of the background
 radiation. Removing from Model~I photoionization by line-of-sight
 quasar leads to Model~II which is represented by the short-dashed
 dotted line. }
\end{figure}

In equation~5, the quasar redshift as seen from the absorbing gas at
$z$ is $z^{\prime}_Q = (1+z_Q)/(1+z)\,-1 $ while the Hubble constant
is $H^{\prime} = H_0
\sqrt{\Omega_{\Lambda}+\Omega_{M}(1+z)^3}$ for a flat universe.  We
assumed the popular $\Lambda$CDM
universe with $\Omega_{\Lambda} = 0.7$ and $\Omega_{M}=0.3$. To
compute the quasar luminosity distance $r_L(z^{\prime}_Q)$, we used
the parametric fit of Pen (1999). The light travel distance used to
derive $dl/dz$ was obtained by integration of the Freedmann
equation. The adopted value for the Hubble constant is $H_0 =
67\,$km/s/Mpc.

\section{Results}

Producing a break near 1050\,\AA\ followed by a steepened power-law of
index $\sim -2$ is fairly easy to achieve if one disregards any
evolution of the background radiation (achieved by setting $\Gamma = 1$ in
Eqn.~2) and adopt the parameters $\xi = -3.5$, $r_{1/2}^0  =
3200$\,Mpc and $n_{H_0}^0 = 1.9 \times 10^{-11}\, {\rm cm^{-2}}$.  This zeroth order
model is illustrated by the short-dashed line in Fig.~1
(Model~I) and is found to be relatively successful in producing a
flat soft ISED. We should point out that photoionization by the
line-of-sight quasar, as represented by the parameter $r_{1/2}^0$, is an
essential ingredient to any successful fit. Actually, if we remove
the $\omega$ term, we consistently obtain a sharp discontinuity at
Ly$\alpha$, as illustrated  
by the short-dashed dotted line of Model~II in Fig.~1.
Including line-of-sight quasar photoionization smooths out completely
the discontinuity provided $r_{1/2}^0$ is sufficiently large. 

\begin{table}
\caption{Model parameters}  
\begin{center}
\begin{tabular}{lrrrc}
\tableline
\tableline
Model & $n^0_{H^0}$  & $r^0_{1/2}$ & $\xi$ & $\Gamma$ evol. \\
(ID) & ($cm^{-3}$) & (Mpc) & --  & (Yes/No)\\
\tableline
I    & $1.9 \times 10^{-11}$ & 3200 &  $-3.5$ & N  \\
II   & $1.9 \times 10^{-11}$ & 0    & $-3.5$ & N \\
III  & $7.4 \times 10^{-11}$ & 3200 &  $-5.0$ & Y  \\
IV   & $4.3 \times 10^{-11}$ & 3200 &  $-4.0$ & Y \\
V    & $3.2 \times 10^{-11}$ & 3000 &  $-3.0$ & Y  \\
VI   & $2.4 \times 10^{-11}$ & 2900 &  $-2.0$ & Y  \\
\tableline

\end{tabular}
\end{center}
\end{table}

\begin{figure}
\plotone{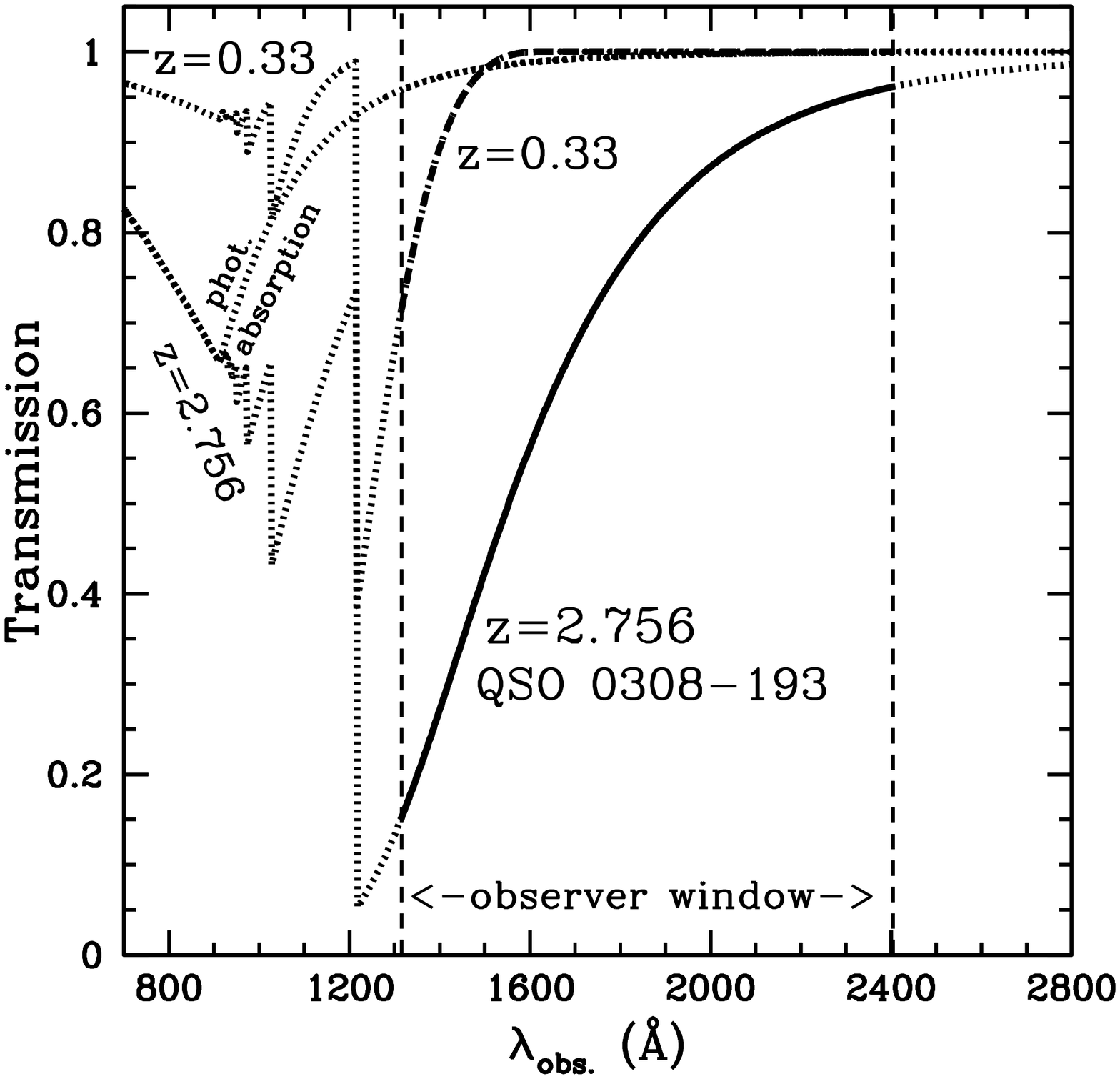}
\caption{The transmission curve as a function of observed wavelength for
our best fit Model~IV in the case of the quasar 0308$-$193 at $z=
2.756$ (solid line). Outside the FOS spectral window (1315--2400\AA),
the dotted line shows the continuation of the calculated transmission
curve. Starting from the right, the first, second and third descending
ramps are contributed in large part by the opacity of Ly$\alpha$,
Ly$\beta$ and Ly$\gamma$ lines, respectively. The small contribution
of photoionization to the opacity is shown by the dotted line labeled
{\it phot. absorption}.  The short-dashed line represents the
transmission curve for the case of a lower redshift quasar at $z= 0.33$. }
\end{figure}

Model~I presents, however, the following deficiency. We have
compelling evidence that the background radiation increases strongly
with $z$ (c.f. Shull et al. 1999; Dav\'e \& Tripp 2001) and references
therein. For this reason, varying $\Gamma(z)$ in equation~2 using the
function proposed by Haardt \& Madau (1996) is preferable and
certainly justified. Models III--VI are such models which span the
range $-$5 to $-$2 in $\xi$. In these models, the parameters
$n_{H_0}^0$ and $r_{1/2}^0$ were varied in such a way that the
different curves superposed as much as possible one another within the
interval 500--1000\,\AA.  All of our model parameters are listed in
Table~1.  We consider Model~IV (solid line) to be the one providing
the best overall fit.  Notice that a broad hump near 500\,\AA\ is
apparent in all four models which consider $J_{\nu_0}^{BCK}$
evolution.  The hump is only present in the case of the ISED derived
from the {\it radio-quiet} subsample, which is the one plotted in
Fig.~1. This hump is more apparent than in the ZKTGD paper because of
our slightly lower wavelength limit at the UV end.  To ascertain the
reality of the hump would require a more extensive sample than the current.

\section{discussion}

In appearance successful, our alternative explanation of the steepening 
of the mean quasar ISED cannot be sustained for the following reasons:

\begin{itemize}

\item {\bf A -- ~} In our calculations, we have treated   $r_{1/2}^0$ as a free
parameter. Therefore, although we adopted a valid prescription for the
variation of the background with $z$ (eq.~3), we have not considered
the issue of the absolute intensity of the background with respect to
the observed quasar intensities. It is straightforward to estimate
$\omega = (F_{\nu}/4\pi J_{\nu_0}^{BCK})^2$ at $z=0$ if we adopt an
observed quasar flux of $F_{\lambda_0}^{Q,0} \sim 2 \times 10^{-14}\,
{\rm erg cm^{-2} s^{-1} }$ near 1000\,\AA, which is typical of the
quasar sample of ZKTGD. The ionization rate coefficient of Haardt \&
Madau (1996) of $4 \times 10^{-14}\, {\rm s^{-1}}$ at $z=0$ translates
into a mean intensity of $J_{\nu_0}^{BCK,0} \approx 1.3 \times^{-23}
\, {\rm erg s^{-1} sterad^{-1} cm^{-2} Hz^{-1}}$, hence $\omega
\approx 5.5 \times 10^{-27}/1.6 \times 10^{-22} = 3.3 \times 10^{-5}$ (using
$F_{\nu_0} = 912^2 \times 3.3 \times 10^{-19} F_{\lambda_0}$).  This
contradicts our requirement that the proximity effect must extends up
to $z=0$ for all quasars whose distance is less than 3\,Gpc (the
approximate value of $r_{1/2}^0$). This estimate of the local value of
$\omega$ clearly shows that the background mean intensity at low $z$
is much stronger than that of the quasar flux reaching us. Another way
of illustrating this is to compare our values of $r_{1/2}$ with the
high redshift value empirically determined of $r_{1/2} \sim 8$\,Mpc by
Bajtlik et~al. (1988). If we translate this estimate to $z=0$ by
taking into account the variation of $J_{\nu_0}^{BCK}(z)$, we obtain
that the Bajtlik et al. local value is $r_{1/2}^0 \approx \sqrt{40}
\times 8 = 50$\,Mpc which is sixty times smaller
than the value of 3200\,Mpc proposed in Table~1. This discrepancy in
the values of $r_{1/2}^0$ (which relates the quasar intensity to the
relative strength of $J_{\nu_0}^{BCK,0}$) effectively rules out our
simple model.

\item {\bf B -- ~} As apparent in Fig.~1, our synthetic 
ISED rolls down steeply at the lower wavelength limit. If we
calculate a similar transmitted ISED but for a reduced range in quasar
redshifts (reducing the highest redshift to 1, say), the roll over
would be shifted to higher wavelength, a feature not seen in any
individual quasar spectrum.

\item {\bf C -- ~} The  self-imposed requirement   that individual absorbers or
clumps be of columns $N_{H^0} \ll 10^{12}\, {\rm cm^{-2}}$ 
implies the existence
of quite a large number of clumps along the line-of-sight. For a column of
$10^{10}\, {\rm cm^{-2}}$, about $6 \times 10^7$ clumps per unit redshift would be
required with an approximate individual mass of about one solar masses
(assuming a neutral hydrogen fraction of $10^{-3}$ and a clump
hydrogen density of order $2 \times 10^{-4}/$cc).

\end{itemize}

In the case of point C, the number of clumps and their masses is not
an unsurmountable objection, however, given the level of speculation
characterizing current ideas about the nature of the dark
matter. Point B above can also be resolved by having the proposed
absorption component disappear at current Epoch (as a result of
Compton heating to $T \gg 10^6\,$K, say).  The strongest objection is
given in which definitely rules out the interplay between quasar and
IGM photoionization as the correct physical mechanism for causing the
smoothing out of the expected sharp break near 1000\AA\ (as the one
seen in Model II of Fig.~1). The next step  might consist in looking for a
different distribution of neutral H than that given in Eqn.~2 which was based
on photoionization of line-of-sight clumps. However, objections could
then arise about justifying the complexity of the function found and
its overall plausibility.

\acknowledgments
This work was supported by the Mexican science funding agency CONACyT
under grant 32139-E and is based on archived data at the Space
Telescope Science Institute which is operated by the Association of
Universities for Research in Astronomy, Inc., under NASA contract
NAS5-26555.  We are indebted to W. Zheng for sharing the original data
presented in ZKTGD.

\end{document}